\documentclass[supp]{new_tlp}
\usepackage{mathptmx}

\usepackage[all]{xy}
\usepackage{theorem}
\usepackage{amssymb}
\usepackage{amsfonts,amsmath,latexsym,amssymb}
\usepackage{tikz}
\usepackage{verbatim}
\usepackage{listings}
\usepackage{color}
\usepackage{multirow}
\usepackage{fancyvrb}
\usepackage{supp}

\newcommand\bcmdtab{\noindent\bgroup\tabcolsep=0pt%
  \begin{tabular}{@{}p{10pc}@{}p{20pc}@{}}}
\newcommand\ecmdtab{\end{tabular}\egroup}

  \title[(Co)recursion in Logic Programming: Lazy vs Eager]{(Co)recursion in Logic Programming:  Lazy vs Eager\footnote{The work of the first two authors was supported by EPSRC Grant EP/K031864/1.}}

   \author[J. Heras, E.~Komendantskaya, and M.~Schmidt]
          {J\'ONATHAN HERAS\\
          School of Computing, University of Dundee, UK\\
          \email{jonathanheras@computing.dundee.ac.uk}\and 
          EKATERINA KOMENDANTSKAYA\\
          School of Computing, University of Dundee, UK\\
          \email{katya@computing.dundee.ac.uk} 
          \and MARTIN SCHMIDT\\
          Institute of Cognitive Science, University of Osnabr\"uck, Germany\\
          \email{martisch@uos.de} 
          }

\pubauthor{Heras, Komendantskaya, and Schmidt}
\jdate{}
\pubyear{}
\pagerange{\pageref{firstpage}--\pageref{lastpage}}
\doi{S1471068401001193}

\newtheorem{example}{Example}[section]
\newtheorem{definition}{Definition}[section]

\definecolor{LightGray}{gray}{0.8}

\begin{document}

\label{firstpage}

\maketitle

  \begin{abstract}
CoAlgebraic Logic Programming (CoALP) is a dialect of Logic Programming designed to bring a more precise compile-time and run-time analysis of 
termination and productivity for recursive and corecursive functions in Logic Programming. 
Its second goal is to introduce guarded lazy (co)recursion akin to functional theorem provers into logic programming. In this paper, we explain 
lazy features of CoALP, and  compare them with the loop-analysis and eager execution in Coinductive Logic Programming (CoLP). We conclude by outlining the future directions in developing the guarded (co)recursion in logic programming. 
  \end{abstract}


\section{Introduction}\label{sec:intro}

Logic Programming (LP) was conceived as a \emph{recursive} programming language for first-order logic. 
Prolog and various other implementations of LP feature \emph{eager} derivations,
and therefore \emph{termination} has been central for logic programming~\cite{deSchreye1994199}.
However, unlike e.g. functional languages, LP has not developed an operational semantics supporting explicit analysis of termination. In typed programming languages like Coq or Agda, 
it is possible to introduce syntactic (static) checks that 
ensure \emph{structural }recursion, and hence termination of programs at run-time. In Prolog, there is no support of this kind.

\begin{example}[BitList]\label{ex:bl}
Consider the following recursive program that defines lists of bits.
\begin{eqnarray*}
1. \texttt{bit(0)} & \gets & \\
2. \texttt{bit(1)} & \gets & \\
3. \texttt{bitlist([])} & \gets & \\
4. \texttt{bitlist([X|Y])} & \gets & \texttt{bit(X)},\texttt{bitlist(Y)}
\end{eqnarray*}
It is  a terminating program, however, if  the order of clauses (3) and (4), or the order of atoms in clause (4) is accidentally swapped, the program would run into an infinite loop. 
\end{example}

This example illustrates that non-terminating (co)recursion is distinguished only empirically at run-time in LP. 
This distinction is not always accurate, and may depend on searching strategies of the compiler,
rather than semantic meaning of the program. 

Coinductive Logic Programming (CoLP)~\cite{GuptaBMSM07,SimonBMG07} has been introduced as a means of supporting \emph{corecursion} in LP. A representative example of coinductive programming is to reason about an infinite data structure, for example an infinite stream of bits. 
	
\begin{example}[BitStream]\label{ex:BS}
Given the definition of bits as in Example~\ref{ex:bl}, an infinite stream of bits is defined as:
    \begin{eqnarray*}
1. \texttt{stream([X|Y])} & \gets & \texttt{bit(X)},\texttt{stream(Y)}
\end{eqnarray*}
Note that unlike \textbf{BitList}, we no longer have the base case for recursion on \texttt{stream}. 
\end{example}

\begin{figure}[t]
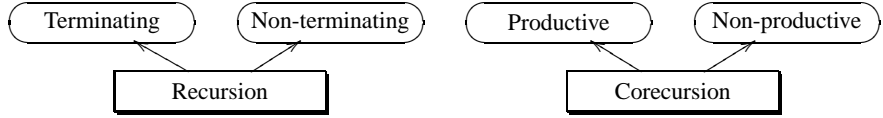

$$
\xy0;/r.12pc/: 
*[o]=<80pt,15pt>\hbox{\txt{Recursion}}="a"*\frm<8pt>{-,},
(-30,18)*[o]=<70pt,15pt>\hbox{\txt{Terminating}}="r"*\frm<8pt>{-},
(30,18)*[o]=<70pt,15pt>\hbox{\txt{Non-terminating}}="b"*\frm<8pt>{-},
(120,0)*[o]=<80pt,15pt>\hbox{\txt{Corecursion}}="c"*\frm<8pt>{-,},
(90,18)*[o]=<70pt,15pt>\hbox{\txt{Productive}}="d"*\frm<8pt>{-},
(150,18)*[o]=<70pt,15pt>\hbox{\txt{Non-productive}}="e"*\frm<8pt>{-},
"a";"b" **\dir{-} ?>*\dir{>},
"a";"r" **\dir{-} ?>*\dir{>},
"c";"d" **\dir{-} ?>*\dir{>},
"c";"e" **\dir{-} ?>*\dir{>},
\endxy
$$
\caption{\footnotesize{\emph{Distinguishing well-founded and non-well-founded cases of recursion and corecursion.}}}\label{pic:sem}
\end{figure}


	The tradition~\cite{Coq94} has a dual notion to termination for well-behaving \emph{corecursion} -- and that is \emph{productivity}. 
 	If termination imposes the condition that any call to an inductively defined predicate like \texttt{bit} must terminate,
 	then productivity requires that every call to a coinductive predicate like \texttt{stream} must \emph{produce} some partially \emph{observed} structure in a finite number of steps. E.g. calling \texttt{stream(X)?}, the program
 	must compute an answer \texttt{[0|Y]} observing the component \texttt{0} in finite time. Moreover, the productivity imposes a second condition:  the computation must be able to proceed corecursively, e.g. in our example, the condition is for 
 	\texttt{Y} to be an infinite productive datastructure. This situation is explained in  e.g.~\cite{AbelPTS13,BK08}.
%
%



CoLP deals with programs like \textbf{BitStream} by using a combination of eager evaluation, SLD-resolution and loop analysis. In simplified terms, for a goal \texttt{stream(X)?} the resolvent loop detection would allow to return an answer
\texttt{X=[0|X]}; by observing the \emph{``regular''} pattern in resolvents involving Clause (1) in the derivations. Similarly to standard (recursive) LP, non-terminating cases of corecursion
(where no regular loop can be found)
 are not formally analysed in CoLP. 
	
	\begin{example}[BadStream]
	\textbf{BadStream} is not productive; that is, it would be executed infinitely without actually constructing a stream:
    \begin{eqnarray*}
1. \texttt{badstream([X|Y])} & \gets & \texttt{badstream([X|Y])}
\end{eqnarray*}
 \end{example}
	
A different case  of corecursion is the below example, which is productive, but cannot be handled by CoLP loop detector, as 	
the stream it defines is not regular. 
	
\begin{example}[TakeFirstN]\label{ex:fs}
The program \texttt{TakeFirstN} defines the stream of natural numbers, and allows to construct a list with the first $n$ elements of the stream by calling \texttt{taken}. 
\begin{eqnarray*}
1. \texttt{from(X,[X|Y])} & \gets & \texttt{from(s(X),Y)}\\
2. \texttt{take(0,Y,[])} & \gets & \\
3. \texttt{take(s(X),[Y|Z],[Y|R])} & \gets & \texttt{take(X,Z,R)}\\
4. \texttt{taken(N,X)} & \gets&  \texttt{from(0,Y)}, \texttt{take(N,Y,X)}
\end{eqnarray*}
\end{example}

In CoLP, calls to e.g. \texttt{taken(s(s(0)),X)?} fall into infinite computations that are not handled by the loop detection procedure.
Similar to how LP would be unable to handle \textbf{BitList} with swapped atoms in clause (4) though in principle the program describes a well-founded inductive structure, 
CoLP would not be able to handle \textbf{TakeFirstN} although it is a perfectly productive stream.
For the query \texttt{taken}, it is intuitively clear that, the construction of the first $n$ elements of the stream should take a finite number of derivation steps. 

Coalgebraic Logic Programming (CoALP)~\cite{KP11-2,KPS12-2} gives a new (coalgebraic) operational semantics for LP; and in particular it offers new methods to analyse  termination and productivity of logic programs.
Using CoALP, we present here a coherent operational treatment of recursion and corecursion in LP, and discuss new methods to distinguish
well-founded and non-well-founded cases of (co)recursion in LP, as outlined in Figure~\ref{pic:sem}.
Unlike Prolog or CoLP, CoALP is a first lazy dialect of logic programming; and it features guarded (co)recursion akin to structural recursion and guarded corecursion in e.g. Coq or Agda~\cite{Coq94,AbelPTS13}.
The current 
implementation of  CoALP in parallel language Go is available in~\cite{KKSH14}; and is tested on a few benchmarks in this paper. 
Here, we abstract from some of the technical details available in~\cite{KPS12-2} and from implementation details available in~\cite{KSH14} and 
give a higher-level discussion of the issues of termination and productivity in LP.

The rest of the paper is structured as follows.
In Section~\ref{sec:cotrees}, we explain the role of laziness in semantics and implementation of CoALP; in Section~\ref{sec:guardedness}, we discuss the effect of guarded corecursion. 
Section~\ref{sec:soundness} is devoted to discussion of our current work on soundness properties for corecursive logic programming. 

\section{Lazy Corecursion in Logic Programming}\label{sec:cotrees}

CoALP uses the standard syntax of Horn-clause logic programming~\cite{Llo88}, but offers a new derivation algorithm in place of the SLD-resolution. 
One of the main distinguishing features of CoALP is its laziness. To our knowledge, it is the first lazy dialect of logic programming.
The issue is best explained using the following example:

\begin{example}\label{ex:SLD}
Given the program \textbf{BitList} and the query \texttt{bitlist([X|Y])}, the standard algorithm of SLD-resolution~\cite{Llo88}  will eagerly attempt to find a derivation, e.g.:
$$\texttt{bitlist([X|Y])} \stackrel{}{\longrightarrow} \texttt{bit(X)},\texttt{bitlist(Y)}  \stackrel{X=0}{\longrightarrow} \texttt{bitlist(Y)}   \stackrel{Y=[]}{\longrightarrow} \Box$$ 

For the program \textbf{BitStream} this will give rise to an infinite SLD-derivation: 
$$\texttt{stream([X|Y])} \stackrel{}{\longrightarrow} \texttt{bit(X)},\texttt{stream(Y)}  \stackrel{X=0}{\longrightarrow} \texttt{stream(Y)}   \stackrel{Y=[X1|Y1]}{\longrightarrow} \texttt{stream([X1|Y1])} \ldots $$

\end{example}

In the above setting, there is no natural place for laziness, as ultimately the strong side of the procedure is a fully automated proof search. Fibrational coalgebraic operational 
semantics of LP presented in~\cite{KPS12-2} inspired us to introduce a structure which we call \emph{coinductive tree}; we use it as a measure for the size of lazy steps in derivations: 
  


\begin{definition}\label{df:coindt}
Let $P$ be a logic program and $G=\gets A$ be an atomic goal.
The \emph{coinductive tree} for $A$ is
  a (possibly infinite) tree $T$ satisfying the following properties.
\begin{itemize}
\item $A$ is the root of $T$.
\item Each node in $T$ is either an and-node (labelled by an atom) or an or-node (labelled by ``$\bullet$''). The root of the tree is an and-node.
\item For every and-node $A'$ occurring in $T$, if there exist exactly $m>0$ 
distinct  clauses $C_1, \ldots , C_m$ in $P$ (a clause $C_i$ has the form $B_i
  \gets B^i_1, \ldots , B^i_{n_i}$, for some $n_i$),  such that $A' = B_1\theta_1 =
  ... = B_m\theta_m$, for mgus $\theta_1, \ldots , \theta_m$,  then $A'$ has exactly $m$ children given by
  or-nodes, such that, for every $i \in \{1, \ldots , m \}$, the $i$th or-node has $n_i$
  children given by and-nodes $B^i_1\theta_i, \ldots ,B^i_{n_i}\theta_i$.

In such a case, we say $C_i$ and $\theta_i$ are \emph{internal resolvents} of $A'$.
\end{itemize}

\end{definition}

Coinductive trees resemble \emph{an-or parallel trees}~\cite{GuptaC94}, see ~\cite{KPS12-2,KSH14} for discussion of their parallel features.
However, they restrict mgus used to form nodes to term-matching. Given two first order atomic formulae $A$ and $B$, an mgu $\theta$ for $A$ and $B$ is called a term-matcher if $A = B\theta$.  
%
In Definition~\ref{df:coindt}, note the condition $A' = B_1\theta_1 = \ldots = B_m\theta_m$. 

\begin{example}
Figure~\ref{pic:stream} shows coinductive trees for various goals in \textbf{BitStream} and \textbf{BitList}; compare with SLD-derivations in Example~\ref{ex:SLD}.
Note that each of those trees is finite by construction of Definition~\ref{df:coindt}; and we do not impose any additional conditions. 
The size of coinductive trees varies, but it is automatically determined by construction of the definition. 
\end{example}
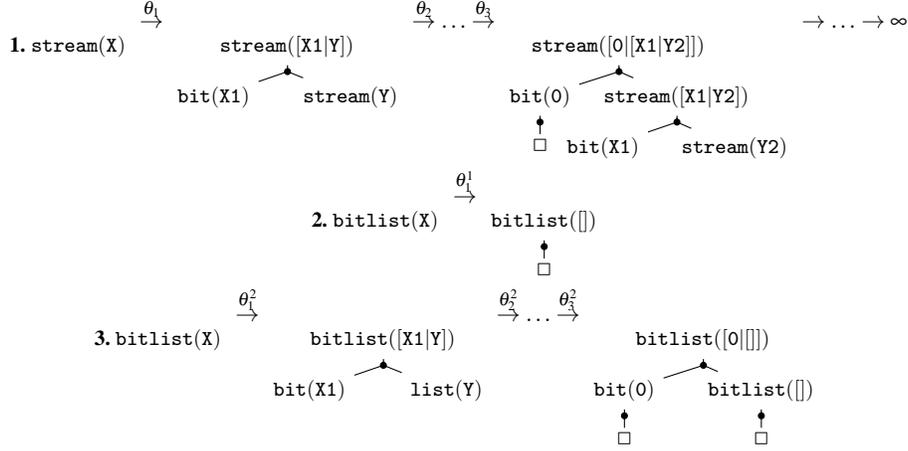
\begin{figure}

  \begin{tikzpicture}[scale=0.28,baseline=(current bounding box.north),grow=down,level distance=12mm,sibling distance=50mm,font=\footnotesize]
  \node {\textbf{1.} $\mathtt{stream(X)}$}; 
  \end{tikzpicture}
$\stackrel{\theta_1}{\rightarrow}$
\begin{tikzpicture}[scale=0.28,baseline=(current bounding box.north),grow=down,level distance=12mm,sibling distance=60mm,font=\footnotesize ]
  \node { $\mathtt{stream([X1|Y])}$}
   child {[fill] circle (4pt)
     child [sibling distance=70mm]{ node {$\mathtt{bit(X1)}$}}
       child { node {$\mathtt{stream(Y)}$}
            }}; 
  \end{tikzpicture}
	$\stackrel{\theta_2}{\rightarrow} \ldots \stackrel{\theta_3}{\rightarrow} $
  \begin{tikzpicture}[scale=0.28,baseline=(current bounding box.north),grow=down,level distance=12mm,sibling distance=35mm,font=\footnotesize]
  \node {$\mathtt{stream([0|[X1|Y2]])}$}
   child {[fill] circle (4pt)
     child [sibling distance=75mm]{ node {$\mathtt{bit(0)}$}
                 child {[fill] circle (4pt)
               child { node {$\Box$}}}}
      child [sibling distance=55mm] { node {$\mathtt{stream([X1|Y2])}$}
           child {[fill] circle (4pt)
     child [sibling distance=70mm]{ node {$\mathtt{bit(X1)}$}}
       child { node {$\mathtt{stream(Y2)}$}
            }}
															}}; 
  \end{tikzpicture}
	$\rightarrow \ldots \rightarrow \infty$

  \begin{tikzpicture}[scale=0.28,baseline=(current bounding box.north),grow=down,level distance=12mm,sibling distance=50mm,font=\footnotesize]
  \node {\textbf{2.} $\mathtt{bitlist(X)}$}; 
  \end{tikzpicture}
$\stackrel{\theta^1_1}{\rightarrow}$
	  \begin{tikzpicture}[scale=0.28,baseline=(current bounding box.north),grow=down,level distance=12mm,sibling distance=50mm,font=\footnotesize]
  \node {$\mathtt{bitlist([])}$}
	child {[fill] circle (4pt)
               child { node {$\Box$}}}; 
  \end{tikzpicture}

  \begin{tikzpicture}[scale=0.28,baseline=(current bounding box.north),grow=down,level distance=12mm,sibling distance=50mm,font=\footnotesize]
  \node {\textbf{3.} $\mathtt{bitlist(X)}$}; 
  \end{tikzpicture}
$\stackrel{\theta^2_1}{\rightarrow}$
\begin{tikzpicture}[scale=0.28,baseline=(current bounding box.north),grow=down,level distance=12mm,sibling distance=60mm,font=\footnotesize ]
  \node {$\mathtt{bitlist([X1|Y])}$}
   child {[fill] circle (4pt)
     child [sibling distance=70mm]{ node {$\mathtt{bit(X1)}$}}
       child { node {$\mathtt{list(Y)}$}
            }}; 
  \end{tikzpicture}
	$\stackrel{\theta^2_2}{\rightarrow} \ldots \stackrel{\theta^2_3}{\rightarrow} $
  \begin{tikzpicture}[scale=0.28,baseline=(current bounding box.north),grow=down,level distance=12mm,sibling distance=35mm,font=\footnotesize]
  \node {$\mathtt{bitlist([0|[]])}$}
   child {[fill] circle (4pt)
     child [sibling distance=75mm]{ node {$\mathtt{bit(0)}$}
                 child {[fill] circle (4pt)
               child { node {$\Box$}}}}
      child [sibling distance=55mm] { node {$\mathtt{bitlist([])}$}
                       child {[fill] circle (4pt)
                child { node {$\Box$}}}}}; 
  \end{tikzpicture}

	\caption{\footnotesize{\emph{\textbf{1:}
Three coinductive trees representing a coinductive derivation for the goal $G = \texttt{stream(X)}$ and the program \textbf{BitStream}, with $\theta_1 = \mathtt{X/[X1|Y]}$, 
$\theta_2 = \mathtt{X1/0}$ and $\theta_3 =\mathtt{Y/[X1|Y2]}$.
\textbf{2-3:} Coinductive trees representing two coinductive derivations for the goal $G = \texttt{bitlist(X)}$ and the program \textbf{BitList}, with $\theta^1_1 =\mathtt{X/[]}$, $\theta^2_1 = \mathtt{X/[X1|Y]}$, 
$\theta^2_2 = \mathtt{X1/0}$, and $\theta^2_3 =\mathtt{Y/[]}$.
}}}
\label{pic:stream} 
\end{figure}

We now define derivations between coinductive trees -- a lazy analogue of SLD-derivations.

\begin{definition}\label{df:coind-res2}
  Let $G= \langle A, T\rangle$ be a goal given by an atom $\gets A$ and the 
  coinductive tree $T$ induced by $A$, and let $C$ be a clause $H \gets
  B_1, \ldots , B_n$.  Then, the goal $G'$ is \emph{coinductively derived}
  from $G$ and $C$ using the mgu $\theta$ if the following conditions
  hold:
\begin{enumerate}
\item[$\bigstar$] $Q(\bar{t})$ is a node in $T$. 
 \item[$\bigstar\bigstar$] $\theta$ is an \emph{mgu} of $Q(\bar{t})$ and $H$.
\item[$\bigstar\bigstar\bigstar$] $G'$ is given by the 
  (coinductive) tree $T\theta$ with the root $A\theta$.
\end{enumerate}

\end{definition}
	
	\begin{definition}\label{df:coind-res3}
A \emph{coinductive derivation} of $P\cup \{G\}$ consists of a
sequence of goals $G= G_0, G_1, \ldots$ 
and a sequence
$\theta_1,\theta_2,\ldots$ of mgus such that each $G_{i+1}$ is
derived from a node $A \in T_i$ (where $T_i$ is the coinductive tree of $G_i$) 
and a clause $C$ using a non-empty substitution $\theta_{i+1}$.
In this case, $\langle A, C, \theta_{i+1}\rangle$ is called a \emph{resolvent}.
\end{definition}
	

Coinductive derivations resemble \emph{tree rewriting}. 
They produce the ``lazy'' corecursive effect:  derivations are given by potentially infinite number of steps, where each individual step is executed in finite time.

\begin{example}
Figure~\ref{pic:stream} shows three possible coinductive derivations for \textbf{BitStream} and \textbf{BitList}. 
Note that two  derivations for \textbf{BitList} terminate (with $\Box$ closing all branches). Note also, that this time, due to the and-or parallel nature of coinductive treees, 
changing the order of atoms or clauses in the program \textbf{BitList} will not change the result. 
\end{example}

 For terminating coinductive derivations, we require at least one or-subtree of the coinductive tree to be closed (with $\Box$ leaves). We also say in such cases that the coinductive tree contains a \emph{success subtree}. The last coinductive trees in the second and third derivation of Figure~\ref{pic:stream} are themselves success subtrees.

Due to its and-or parallel properties~\cite{KSH14}, CoALP is more robust  than eager sequential SLD-resolution when it comes to reflecting program's operational meaning; and mere change in 
the clause order would not 
place a terminating recursive function into a non-terminating class, cf.~Figure~\ref{pic:sem}. 
Yet more importantly, this new coinductive derivation procedure allows us to characterise productive and non-productive programs with better precision.
In Introduction, we have seen that according to eager interpreter of CoLP, both programs \textbf{BadStream} and \textbf{TakeFirstN} are non-terminating; despite of one being productive, and another -- non-productive.
Next example shows that under lazy execution, productive programs with irregular pattern of resolvents can be handled more naturally.



\begin{figure}
\centering
 \begin{tikzpicture}[level 1/.style={sibling distance=15mm},
level 2/.style={sibling distance=20mm},
level 3/.style={sibling distance=20mm},scale=.8,font=\scriptsize]
  \node (root) {\texttt{taken($s^2$(0),X)}}[level distance=6mm]
                     child { [fill] circle (1pt)
                             child { node {\texttt{from(0,Y)}}}
                             child { node {\texttt{take($s^2$(0),Y,X)}} }
           };
  \end{tikzpicture}
  \begin{tikzpicture}
     \draw (0,2) node{{\scriptsize $\xrightarrow{\theta_1}$}};
     \draw (0,0) node{\textcolor{white}{$\rightarrow$}};
  \end{tikzpicture}
\hspace{-1cm}
 \begin{tikzpicture}[level 1/.style={sibling distance=15mm},
level 2/.style={sibling distance=30mm},
level 3/.style={sibling distance=20mm},
level 5/.style={sibling distance=20mm},scale=.8,font=\tiny]
  \node (root) {\texttt{taken($s^2$(0),[X1|R1])}}[level distance=6mm]
                     child { [fill] circle (1pt)
                             child { node {\texttt{from(0,[X1|Y1])}}}
                             child { node {\texttt{take($s^2$(0),[X1|Y1],[X1|R1])}} 
                                     child { [fill] circle (1pt)
                                              child { node {\texttt{take(s(0),Y1,R1)}}}
                                           }
                                    }
           };
  \end{tikzpicture}
  \begin{tikzpicture}
     \draw (0,3) node{{\scriptsize $\xrightarrow{\theta_2}$}};
     \draw (0,0) node{\textcolor{white}{$\rightarrow$}};
  \end{tikzpicture}
\hspace{-1cm}
 \begin{tikzpicture}[level 1/.style={sibling distance=15mm},
level 2/.style={sibling distance=35mm},
level 3/.style={sibling distance=20mm},scale=.8,font=\tiny]
\node (root) {\texttt{taken($s^2$(0),[X1,X2|R2])}}[level distance=6mm]
                     child { [fill] circle (1pt)
                             child { node {\texttt{from(0,[X1, X2|Y2])}}}
                             child { node {\texttt{take($s^2$(0),[X1,X2|Y2],[X1,X2|R2])}} 
                                     child { [fill] circle (1pt)
                                              child { node {\texttt{take(s(0),[X2|Y2],[X2|R2])}}
                                                      child { [fill] circle (1pt)
                                                              child { node {\texttt{take(0,Y2,R2)}}}
                                                            }
                                                    }
                                           }
                                    }
           };
  \end{tikzpicture}
 \caption{\footnotesize{\emph{First steps of the derivation for the goal \texttt{taken($s^2$(0),X)} -- \texttt{$s^2$(0)} denotes \texttt{s(s(0))} -- and the program \texttt{TakeFirstN}, with $\theta_1=\mathtt{Y/[X1|Y1],X/[X1|R1]}$ and $\theta_2=\mathtt{Y1/[X2|Y2], R1/[X2|R2]}$. As \texttt{take} is an inductive predicate, and \texttt{from} is coinductive; resolvents for \texttt{take} nodes are given priority. 
}}}\label{fig:fibs} 
\end{figure}
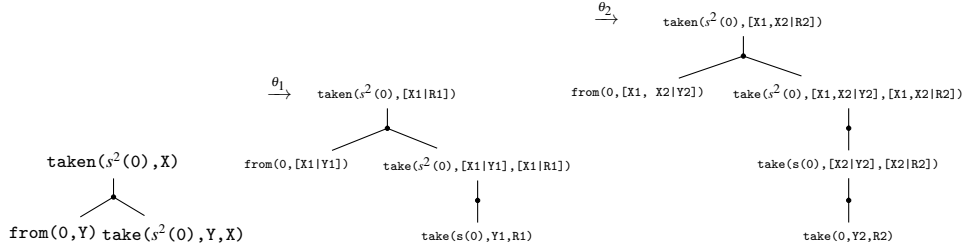

\begin{example}
Figure~\ref{fig:fibs} shows the first steps in the derivation for the program \textbf{TakeFirstN} and the goal \texttt{taken(s(s(0)),X)}.
Unlike CoLP, CoALP is able to compute the second element of the stream in finite time.
\end{example}

There will be classes of non-terminating and non-productive programs for which coinductive trees grow infinite, and lazy derivations fail being "lazy". The program \textbf{BadStream} 
is one such example. We will consider this issue in the next section.

\section{Guarding (Co)recursion}\label{sec:guardedness}

The previous section introduced coinductive trees, which allowed us to distinguish terminating and productive programs like \textbf{BitStream}, \textbf{BitList}, \textbf{TakeFirstN} from non-productive programs like \textbf{BadStream}, by simply observing that coinductive trees remain finite for the former, while growing infinite for the latter.
It was especially significant that this new approach was, unlike Prolog, robust to permutations of clauses and atoms, and, unlike CoLP, was working with productive irregular streams.  
Curiously, the following logic program fails to produce finite coinductive trees:

\begin{example}[\textbf{GC}]\label{ex:GC}
Let \texttt{GC} (for graph connectivity) denote the logic program
\begin{eqnarray*}
1. \texttt{connected(X,X)} & \gets &\\
2. \texttt{connected(X,Y)} & \gets & \texttt{edge(X,Z)},\texttt{connected(Z,Y)}
\end{eqnarray*}
It would be used with database of graph edges, like $\texttt{edge(0,1)} \gets$.

The program gives rise to infinite coinductive trees, see Figure~\ref{pic:GC}. It would terminate in LP, but, 
similarly to our discussion of \textbf{BitList}, would lose the termination property if the order of clauses (1) and (2) changes, or if the order of the atoms in clause (2) changes.

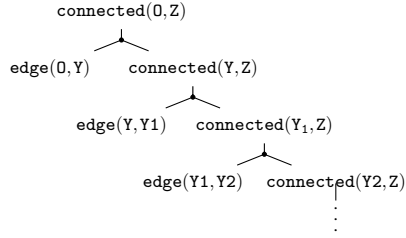
\begin{figure}
\begin{center}
  \begin{tikzpicture}[scale=0.19,baseline=(current bounding box.north),grow=down,level distance=20mm,sibling distance=100mm,font=\scriptsize]
  \node {$\mathtt{connected(O,Z)}$}
   child {[fill] circle (4pt)
     child { node {$\mathtt{edge(O,Y)}$}}
       child { node {$\mathtt{connected(Y,Z)}$}
        child {[fill] circle (4pt)
       child { node{$\mathtt{edge(Y,Y1)}$}}
         child { node{$\mathtt{connected(Y_1,Z)}$}
				        child {[fill] circle (4pt)
       child { node{$\mathtt{edge(Y1,Y2)}$}}
         child { node{$\mathtt{connected(Y2,Z)}$}
				child {node{$ \vdots$}}}}}}}}; 
  \end{tikzpicture}
 \end{center}
\caption{\footnotesize{\emph{The infinite coinductive tree for the program GC from from Example~\ref{ex:GC}, for the database $\texttt{edge(0,1)} \gets$.
} }}
\label{pic:GC} 
\end{figure}

\end{example}

The reason behind infinity of coinductive trees for the above program is the absence of function symbols -- ``constructors" in the clause heads.
The lazy nature of coinductive trees was in part due to the term-matching used to compute them. Term-matching loses its restrictive power in the absence of constructors.
A very similar procedure of guarding recursion by constructors of types is used in e.g. Coq or Agda. This observation would suggest an easy way to fix the \textbf{GC} example, by introducing reducing dummy-constructors:

\begin{example}[\textbf{Guarded GC}]\label{ex:gGC}
\begin{eqnarray*}
1. \texttt{connected(X,cons(Y,Z))} & \gets &  \texttt{edge(X,Y),connected(Y,Z)}\\
2. \texttt{connected(X,nil)} & \gets & \\
\end{eqnarray*}
\end{example}

Considerations of this kind led us to believe that our lazy (co)recursive approach opens the way for a compile-time termination and productivity checks akin to respective checks in Coq or Agda~\cite{Coq94,AbelPTS13}. The programmer would be warned of non-terminating cases and asked to find a guarded reformulation for his functions. In Coq and Agda, different checks are imposed on recursive functions (``structural recursion" condition) and corecursive functions (``guardedness" checks). 
In logic programming terms, where types or predicate annotations are unavailable,  we can formulate a uniform productivity property for recursive and corecursive programs, as follows:


\begin{definition}\label{df:prod}
Let $P$ be a logic program, $P$ is \emph{productive} if for any goal $G$, the coinductive tree for $P\cup \{G\}$ has a  finite size.
\end{definition}

The above is a semantic property; syntactically, we need to introduce  guardedness checks to ensure productivity. The intuitive idea is to ensure that every coinductive program behaves like \texttt{BitStream}: \texttt{BitStream} is guarded by the coinductive function symbol (or ``guard'') \texttt{scons} (denoted by \texttt{[.|.]});
and hence all coinductive trees for it are finite, see Figure~\ref{pic:stream}.
On the contrary, \textbf{Comember} lacks a guarding constructor. 

\begin{example}[Comember]\label{ex:comember}
The predicate \texttt{comember} is true if and only if the element \texttt{X} occurs an infinite number of times in the stream \texttt{S}.
\begin{eqnarray*}
1. \texttt{drop(H,[H|T],T)} & \gets & \\
2. \texttt{drop(H,[H1|T],T1)} & \gets & \texttt{drop(H,T,T1)}\\
3. \texttt{comember(X,S)} & \gets & \texttt{drop(X,S,S1)},\texttt{comember(X,S1)}
\end{eqnarray*}

\texttt{Comember} is un-productive for e.g. the coinductive tree arising from the query \texttt{comember(X,S)} contains a chain of alternating 
$\bullet$'s and atoms \texttt{comember(X,S1)}, \texttt{comember(X,S2)}, etcetera, yielding an infinite coinductive tree.
\end{example}

We will give a high-level formulation of guardedness checks here, for more technical discussion, see \cite{KPS12-2}.

\textbf{Guardedness Check 1 (GC1)}: If the same predicate $Q$ occurs in the head and in the body of a clause, then there must exist a function symbol $f$ occurring among the arguments of $Q$; such that the number of its occurrences is reduced from head to body.

\begin{example}[Guarded Comember]\label{ex:gcomember}
We propose the following guarded definition of comember, thereby simplifying it and reducing an extra argument to \texttt{drop}.
\begin{eqnarray*}
1. \texttt{drop(H,[H|T])} & \gets & \\
2. \texttt{drop(X,[H|T])} & \gets & \texttt{drop(X,T)}\\
3. \texttt{gcomember(X,[H|T])} & \gets &  \texttt{drop(X,[H|T])} , \texttt{gcomember(X,T)}
\end{eqnarray*}
In CoALP, the goal  \texttt{gcomember(0,nats)} will lazily search for  $0$ in an infinite stream of natural numbers, but it never falls into un-productive coinductive trees, as CoLP would do. 
\end{example}

\textbf{GC1} would be sufficient for some programs, like \texttt{BitStream}, where there is only one (co)inductive clause; 
but not in the general case. 
LP in general is not compositional, that is, composing two programs may yield a program that has semantic properties not present in the initial programs.
Same rule applies in CoALP: if both $P_1$ and $P_2$ are productive programs, their composition is not guaranteed to be a productive program; 
the next check is imposed to cover the compositional cases.

\textbf{Guardedness Check 2 (GC2)}: For every clause head $A$, construct a coinductive tree with the root $A$. If there are atoms $Q(\bar{t})$ and $Q(\bar{t'})$ in the coinductive tree such that 
$Q(\bar{t'})$ is a child of $Q(\bar{t})$, apply \textbf{GC1} to the clause   
 $Q(\bar{t}) \gets Q(\bar{t'})$.

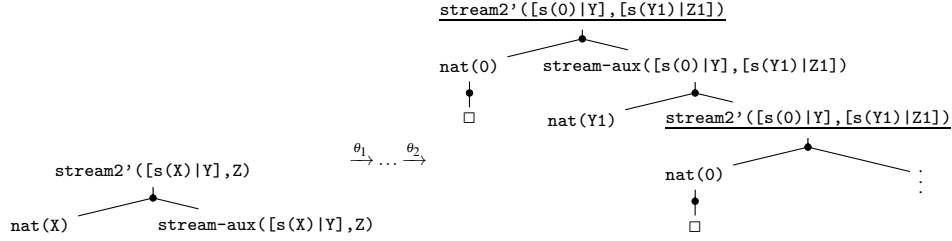
\begin{figure}
\centering
 \begin{tikzpicture}[level 1/.style={sibling distance=30mm},
level 2/.style={sibling distance=50mm},
level 3/.style={sibling distance=50mm},scale=.6,font=\scriptsize]
  \node (root) {\texttt{stream2'([s(X)|Y],Z)}} [level distance=6mm]
     child { [fill] circle (2pt)
             child { node {\texttt{nat(X)}}}
             child { node {\texttt{stream-aux([s(X)|Y],Z)}}}
             };
\end{tikzpicture} 
\hspace{-.7cm}
\begin{tikzpicture}
     \draw (0,1) node{{\scriptsize $\xrightarrow{\theta_1} \ldots \xrightarrow{\theta_2}$}};
     \draw (0,0) node{\textcolor{white}{$\rightarrow$}};
\end{tikzpicture}
\hspace{-.2cm}
\begin{tikzpicture}[level 1/.style={sibling distance=30mm},
level 2/.style={sibling distance=50mm},
level 3/.style={sibling distance=50mm},scale=.6,font=\scriptsize]
  \node (root) {\underline{\texttt{stream2'([s(0)|Y],[s(Y1)|Z1])}}} [level distance=6mm]
     child { [fill] circle (2pt)
             child { node {\texttt{{nat(0)}}}
                          child { [fill] circle (2pt)
                                  child { node {$\Box$}}}}
             child { node {\texttt{stream-aux([s(0)|Y],[s(Y1)|Z1])}}
                         child { [fill] circle (2pt)
                                  child { node {\texttt{nat(Y1)}}}
                                  child { node {\underline{\texttt{stream2'([s(0)|Y],[s(Y1)|Z1])}}}
                                  child { [fill] circle (2pt)
                                           child { node {\texttt{nat(0)}}
                                                   child { [fill] circle (2pt)
                                                    child { node {$\Box$}}}}
                                           child { node {\vdots}}}}
                                          }
                   }
             };
\end{tikzpicture}
\caption{{\emph{\scriptsize Coinductive derivation of \texttt{stream2'([s(X)|Y],Z)} and the program from Example~\ref{ex:streamaux} producing an infinite coinductive tree, with $\theta_1=\texttt{X/0}$ and $\theta_2=\texttt{Z/[s(Y1)|Z1]}$. The figure also represents one GC-derivation generated during \textbf{GC3}. \textbf{GC3} detects the un-guarded loop; see the underlined atoms.}}}\label{fig:unpro} 
\end{figure}

\textbf{GC1--GC2} handle some programs well, but they are  still insufficient in the general case.
The following  program  passes the checks \textbf{GC1-GC2}, but is not productive in the sense of Definition~\ref{df:prod}, see Figure~\ref{fig:unpro}.
 
\begin{example}[Un-productive Program  that passes GC1-GC2]\label{ex:streamaux}
\begin{eqnarray*}
1. \texttt{stream2'([s(X)|Y],Z)} &\gets& \texttt{nat(X), stream-aux([s(X)|Y],Z)}\\
2. \texttt{stream-aux(X,[s(Y)|Z])} &\gets& \texttt{nat(Y), stream2'(X,[s(Y)|Z])}
\end{eqnarray*}
\end{example} 


\textbf{Guardedness Check 3 (GC3)}:
For every clause head $A$, start a coinductive derivation with the goal $A$ imposing \textbf{GC2} condition to every coinductive tree in the derivation, and imposing the following termination conditions:
\begin{enumerate}
	\item Terminate coinductive derivation if \textbf{GC2} fails for at least one tree.
	\item Terminate coinductive derivation if all branches are either closed with $\Box$ or contain guarded loops only.
\end{enumerate}

Note that the checks \textbf{GC1-GC3} we have introduced here are a pre-processing (compile-time) mechanism of CoALP. Once the program passed the guardedness checks, it does coinductive derivations lazily; and does not require any loop-detection procedures at run-time. 
If a program fails \textbf{GC1-GC3}, the programmer will be asked to re-formulate the definitions as we have seen in Examples~\ref{ex:gGC} and \ref{ex:gcomember}. 
The first implementation of guardedness checks is available at~\cite{KKSH14}.

We finish this section with Table~\ref{tab:benchmark} comparing how SWI-Prolog, CoLP and CoALP handle various recursive and corecursive programs. For CoALP, we also benchmark guardedness checks.

For coinductive programs, CoLP can only handle coinductive programs that contain a regular pattern and fails otherwise (cf. Table~\ref{tab:benchmark}); on the contrary, CoALP, in its lazy style, works for any productive program. This is illustrated, for instance, with the programs \texttt{TakeFirstN} and \texttt{TakeRepeat}.
Table~\ref{tab:benchmark} shows that CoALP is slower than the CoLP interpreter and SWI-Prolog -- note that SWI-Prolog is a fully-tuned mature programming language and the CoLP interpreter runs on top of SWI-Prolog, as opposed to our implementation of CoALP in Go from scratch.

\begin{table}[t]
\centering
{\scriptsize 
\begin{tabular}{c|cc|c|c|}

& \multicolumn{2}{c|}{\multirow{2}{*}{CoALP}} & \multirow{2}{*}{CoLP} & \multirow{2}{*}{SWI-Prolog}\\
 &  & & &\\ 
\cline{1-5}
\multicolumn{1}{ c|  }{\multirow{2}{*}{TakeFirstN$\dag$} } & \multirow{2}{*}{Yes} & GC time: $0.0002s$ & \multirow{2}{*}{No} & \multirow{2}{*}{No}\\ \cline{3-3}
\multicolumn{1}{ c|  }{}&  & runtime: lazy execution  &  &     \\ \cline{1-5} 
\multicolumn{1}{ c|  }{\multirow{2}{*}{Takerepeat$\dag$} }  & \multirow{2}{*}{Yes} & GC time: $0.0009s$ & \multirow{2}{*}{Yes ($0.0001s$)} & \multirow{2}{*}{No}\\ \cline{3-3}
\multicolumn{1}{ c|  }{} &  & runtime: lazy execution  &  &     \\ \cline{1-5} 
\multicolumn{1}{ c|  }{\multirow{2}{*}{Comember$\dag$}}  & \multicolumn{2}{c|}{\multirow{2}{*}{Not guarded}}  & \multirow{2}{*}{Yes$^{?}$ ($0.0001s$)} & \multirow{2}{*}{No}\\ 
\multicolumn{1}{ c| }{} &  &   &  &     \\ \cline{1-5} 
\multicolumn{1}{ c|  }{\multirow{2}{*}{GComember$\dag$}}  & \multirow{2}{*}{Yes} & GC time:  $0.0011s$ & \multirow{2}{*}{Yes$^{?}$ ($0.0001s$)} & \multirow{2}{*}{No}\\ \cline{3-3}
\multicolumn{1}{ c| }{} &  & runtime: lazy execution  &  &     \\ \cline{1-5} 
\multicolumn{1}{ c|  }{\multirow{2}{*}{SumFirstn$\dag$}}  & \multirow{2}{*}{Yes} & GC time: $0.0013s$ & \multirow{2}{*}{No} & \multirow{2}{*}{No}\\ \cline{3-3}
\multicolumn{1}{ c| }{} &  & runtime: lazy execution  &  &     \\ \cline{1-5} 
\multicolumn{1}{ c| }{\multirow{2}{*}{\texttt{FibStream}$\dag$} } & \multirow{2}{*}{Yes} & GC time: $0.0006s$ & \multirow{2}{*}{No} & \multirow{2}{*}{No}\\ \cline{3-3}
\multicolumn{1}{ c|  }{} &  & runtime: lazy execution  &  &     \\ \cline{1-5} 
\multicolumn{1}{ c|  }{\multirow{2}{*}{Infinite Automata$\dag$} }  & \multirow{2}{*}{Yes} & GC time: $0.0011s$ & \multirow{2}{*}{Yes ($0.0001s$)} & \multirow{2}{*}{No}\\ \cline{3-3}
\multicolumn{1}{ c|  }{} &  & runtime: lazy execution  &  &     \\ \cline{1-5} 
\multicolumn{1}{ c|  }{\multirow{2}{*}{Knights}}      & \multirow{2}{*}{Yes} & GC time: $0.225s$ & \multirow{2}{*}{Yes ($1.13s$)} & \multirow{2}{*}{Yes ($0.012s$)}\\ \cline{3-3}
\multicolumn{1}{ c| }{}&  & runtime: $3.002s$  &  &     \\ \cline{1-5} 
\multicolumn{1}{ c|  }{\multirow{2}{*}{Finite Automata} }  & \multirow{2}{*}{Yes} & GC time: $0.0011s$ & \multirow{2}{*}{Yes ($0.04s$)} & \multirow{2}{*}{Yes ($0.0005s$)}\\ \cline{3-3}
\multicolumn{1}{ c|  }{}&  & runtime: $0.0023s$  &  &     \\ \cline{1-5} 
\multicolumn{1}{ c|  }{\multirow{2}{*}{Ackermann} }  & \multirow{2}{*}{Yes} & GC time: $0.001s$ & \multirow{2}{*}{Yes ($7.692s$)} & \multirow{2}{*}{Yes ($3.192s$)}\\ \cline{3-3}
\multicolumn{1}{ c|  }{} &  & runtime: $13.23s$&  &     \\ \cline{1-5} 
\end{tabular}
}

\caption{{\footnotesize Execution of different programs in CoALP, CoLP and SWI-Prolog. Examples marked with $\dag$ involve both inductive and coinductive predicates. In the table, ``No'' means that the system runs forever without returning an answer, and ``Yes$^?$'' indicates that the program succeeds if it contains a regular pattern and fails otherwise.  
}}\label{tab:benchmark}
\end{table}

\section{Work-in-Progress: Soundness for Corecursion}\label{sec:soundness}

There are two main directions for CoALP's development, both related to soundness: 

(I)	We are in the process of establishing soundness of \textbf{GC1-GC3} that is, the property that, \emph{if a program $P$ is guarded by \textbf{GC1-GC3}, then it is productive in CoALP}. 
	
	Proving this property in the general case is a challenge; and involves pattern analysis on resolvents and also a proof of termination of \textbf{GC1-GC3}. 
	Example~\ref{ex:streamaux} and Figure~\ref{fig:unpro} give a flavor of the complicated cases the guardedness checks need to cover.
	Note that \textbf{GC1-GC3} provide the guarding property only in the CoALP setting, and the same idea of guarding (co)recursion by constructors would fail for standard LP or CoLP, as many examples of this paper show.
	
	(II) Soundness of coinductive derivations needs to be established. This challenge is best illustrated by the following example. 

\begin{example}[Soundness for Comember]
To check the validity of a query in \texttt{Comember} (Example~\ref{ex:comember}) for an arbitrary stream, one needs to satisfy  two conditions:
1) finding an element to \texttt{drop} in a finite time, 2) finding guarantees that this finite computation will be repeated an infinite number of times for the given stream.
CoLP would handle such a case for all streams that consist of a regular finite repeating pattern and will not be able to handle cases when the input stream is not regular.
CoLP would fail to derive true or falsity of e.g. the query \texttt{comember(0,nats)}, where \texttt{nats} is the stream of natural numbers, as CoLP falls into an infinite non-terminating computation and fails to 
produce any response to the query.
CoALP in its current implementation will handle any case of corecursion, including  \texttt{comember(0,nats)}, but in its lazy, and therefore partial, style. 
\end{example}

Similarly, \texttt{TakeFirstN} falls into an infinite loop with CoLP, but unfolds lazily with CoALP, see Figure~\ref{fig:fibs}. Laziness on its own, however, does not guarantee soundness.  

For inductive programs and recursive functions, CoALP yields the same theorems of soundness and completeness as classical LP~\cite{Llo88}; cf.~\cite{KPS12-2}. The only adaptation to the already described coindutive derivation procedure is the requirement that the derivation terminates and gives an answer whenever a \emph{success subtree} is found. Thus, generalisation of standard soundness and completeness for induction in CoALP is not very surprising.

Soundness of CoALP for coinductive programs is conceptually more interesting: it has to include a number of guarantees that need to be checked at compile-time and run-time, that is:
\begin{enumerate}
	\item Identification of the guarding pattern coming from sound guardedness checks;
	\item Guarantee that the guarding pattern will be produced in a finite number of derivation steps;
	\item Guarantee that the guarding pattern will be re-produced an infinite number of times.
\end{enumerate}

Item 3. in particular may allow for a few different solutions. In its basic form, it can be a repeated regular pattern, as it is done in CoLP. In a more sophisticated form, it can cover irregular patterns, as long as more involved guarantees of infinite execution are be provided, cf. Example~\ref{ex:fs} and Figure~\ref{fig:fibs}.

To conclude, we have described a new method to analyse termination and productivity of logic programs by means of lazy guarded corecursion in CoALP, as outlined in Figure~\ref{pic:sem}. We advocated a new style of programming in LP, where the programmer is in charge of providing termination or productivity measures for (co)recursive programs at compile-time, as it is done in some other declarative languages with recursion and corecursion. Finally, we outlined the main directions towards establishing soundness results for CoALP outputs.

\bibliographystyle{acmtrans}
\bibliography{katya2}






\label{lastpage}
\end{document}